\documentclass[journal]{IEEEtran}

\usepackage[english]{babel}

\usepackage{cite}
\usepackage[utf8]{inputenc}
\usepackage{amsmath}
\usepackage{multirow}
\usepackage{graphicx}
\usepackage[english]{babel}
\usepackage{amsfonts}
\usepackage{amsthm}
\usepackage{subfig}
\usepackage{lmodern}
\usepackage{url}
\newtheorem*{remark}{Remark}

\usepackage[colorlinks=true, allcolors=blue]{hyperref}
\def\BibTeX{{\rm B\kern-.05em{\sc i\kern-.025em b}\kern-.08em
    T\kern-.1667em\lower.7ex\hbox{E}\kern-.125emX}}

\title{Context-Conditioned Spatio-Temporal Predictive Learning for Reliable V2V Channel Prediction 
\thanks{Part of this work was supported by the California Transportation Department and by the National Science Foundation.}}

\author{Lei~Chu,~\IEEEmembership{Senior Member,~IEEE,} Daoud Burghal, Rui Wang, Michael Neuman, 
        and Andreas~F.~Molisch,~\IEEEmembership{Fellow,~IEEE}
}

\begin{document}
\maketitle

\begin{abstract}
Achieving reliable multidimensional Vehicle-to-Vehicle (V2V) channel state information (CSI) prediction is both challenging and crucial for optimizing downstream tasks that depend on instantaneous CSI. This work extends traditional prediction approaches by focusing on four-dimensional (4D) CSI, which includes predictions over time, bandwidth, and antenna (TX and RX) space. Such a comprehensive framework is essential for addressing the dynamic nature of mobility environments within intelligent transportation systems, necessitating the capture of both temporal and spatial dependencies across diverse domains. 
To address this complexity, we propose a novel context-conditioned spatiotemporal predictive learning method. This method leverages causal convolutional long short-term memory (CA-ConvLSTM) to effectively capture dependencies within 4D CSI data, and incorporates context-conditioned attention mechanisms to enhance the efficiency of spatiotemporal memory updates. Additionally, we introduce an adaptive meta-learning scheme tailored for recurrent networks to mitigate the issue of accumulative prediction errors. 
We validate the proposed method through empirical studies conducted across three different geometric configurations and mobility scenarios. Our results demonstrate that the proposed approach outperforms existing state-of-the-art predictive models, achieving superior performance across various geometries. Moreover, we show that the meta-learning framework significantly enhances the performance of recurrent-based predictive models in highly challenging cross-geometry settings, thus highlighting its robustness and adaptability.
 
\end{abstract}


\section{Introduction}

Vehicle-to-vehicle (V2V) communications are crucial for future driving, especially for assisted or autonomous systems \cite{tataria20216g, liu2023self, noor20226g, molisch2024deep}. These systems enable vehicles to warn each other of imminent actions like emergency braking or coordinate smooth lane changes. However, widespread adoption of V2V communication has been slow due to, at least partially, economic factors and the unpredictable performance of V2V systems. The latter can be attributed to challenges that include signal propagation issues and high device density, causing interference and packet loss \cite{tang2019path, molisch2021measurement}. Thus, improving the reliability and latency of V2V links is essential.
Due to the high dynamics in V2V environments, a key challenge is to maintain a robust communication with outdated channel measurements. Therefore, effective channel prediction methods are needed to infer the current state from past data. 

The significance of channel prediction for V2V scenarios is widely acknowledged in the literature \cite{chu2024model, 10286287, jiang2019neural, liao2023machine, bogale2020adaptive}, with numerous papers addressing this topic. Most studies employ classical methods, such as Extended Kalman Filters (e.g., \cite{aghamohammadi1989adaptive, liao2020nonlinear}), or sparsity-based approaches (e.g., \cite{gao2015spatially, groll2019sparsity}). While previous research has shown that these algorithms perform well with theoretical channel models, they face challenges when applied to real-world data. This is due to the mismatch between the underlying models of these classical methods and physical reality, as well as their inability to predict channels over longer timescales.

\section{Related Works}

\subsection{Machine Learning based V2V Channel Prediction}

Machine learning (ML) provides a framework for making decisions and predictions from available data without relying on specific analytical models \cite{mao2018deep, huang2019machine}. This approach has revolutionized the handling of previously insurmountable computational challenges. Consequently, ML-based channel estimation is conjectured to perform better for these purposes, as it can predict channels over larger distances and uncover hidden relationships over time \cite{luo2018channel, yuan2020machine}. ML has been applied in various settings, including channel prediction in massive MIMO \cite{kim2020massive}, high-mobility massive MIMO-OFDM \cite{wu2021channel, qin2022partial, liu2022spatio, ma2018sparse, 10286287}, vehicle-to-infrastructure \cite{bogale2020adaptive}, cross-band channel prediction \cite{bakshi2019fast}, vehicular edge networks \cite{liao2021deep}, and UAV channels \cite{ladosz2019gaussian}.

While these applications are promising, they do not directly address the unique characteristics of V2V channels, whose dominant propagation effects differ fundamentally from those observed in infrastructure-based communications \cite{beygi2015nested}. Although there are some investigations into V2V channel prediction using ML (e.g., \cite{va2016impact, ma2018sparse, liao2023machine}), studies based on real-world data are exceedingly rare. The only directly relevant studies we are aware of are \cite{ramya2019using}, which relies exclusively on path loss measurements, and \cite{joo2019deep}, which extracts CSI from 802.11p on-board units to predict received power. However, these studies have limitations: the units used were not calibrated, and only single-antenna measurements were performed. In contrast, 5G NR V2V systems \cite{garcia2021tutorial, noor20226g} are expected to employ multiple antenna elements. We conjecture that the primary reason for this gap is the scarcity of measurement data available to ML research groups, which hinders the development and application of more accurate models for V2V channels.

\subsection{Predictive Learning Algorithms}

In this work, we address multi-dimensional channel predictions based on multiple V2V measurement campaigns, encompassing a range of scenarios from low mobility (such as campus streets or city canyons) to high mobility (such as highways). The solutions for multi-dimensional channel predictions in the context of deep learning are related to the domain of \textit{spatio-temporal predictive learning}.  In the literature, these methods can be broadly classified into two categories: recurrent-free 
and  recurrent-based predictive learning algorithms \cite{tan2023openstl}. Recurrent-free models perform the prediction by directly feeding the entire sequence of observed frames into the model, which then outputs the complete set of predicted frames all at once \cite{vaswani2017attention, jiang2022accurate, gao2022simvp}. On the other hand, recurrent-based models attempt to make predictions on a frame-by-frame basis. For example, LSTM-based recurrent neural networks (RNN) have been extensively utilized for the modeling and analysis of time series data due to their ability to capture long-term dependencies and manage issues related to vanishing gradients \cite{gers2002learning}. However, LSTM networks are inherently designed to handle one-dimensional sequential data, which limits their effectiveness in applications requiring the integration of both spatial and temporal information. To address this limitation, Shi et al. introduced the ConvLSTM network, a prototypical architecture that extends the conventional LSTM by incorporating convolutional structures within the gating mechanisms \cite{shi2015convolutional}. This advancement enables the ConvLSTM network to effectively model spatial-temporal dependencies, thereby offering a robust solution for complex tasks such as precipitation nowcasting and other applications involving dynamic spatial data. Moreover, it has proven effective in modeling statistical wireless channel dependencies \cite{liu2022spatio, huang2024frequency}. 
Recently, a new spatiotemporal LSTM (ST-ConvLSTM) unit, which simultaneously extracts and memorizes spatial and temporal representations, was introduced in \cite{wang2017predrnn} (extended version in  \cite{wang2022predrnn}). The ST-ConvLSTM has proven to be a state-of-the-art (SOTA) spatio-temporal predictive learning model, achieving SOTA performance across many datasets, as verified in \cite{tan2023openstl}. Due to limited space, interested readers are referred to a recent survey for more related ML models in the literature \cite{tan2023openstl}. 

In summary, each type of method has its own strengths and limitations. Transformers are highly effective at extracting semantic correlations in long sequences; However, in multiple dimensional time series modeling, the goal is to capture temporal relationships within an ordered sequence of continuous points in spatial-temporal domains. Although positional encoding and token embeddings help maintain some ordering, the permutation-invariant self-attention mechanism may lead to a loss of temporal information, as demonstrated in \cite{zeng2023transformers}. On the other hand, ConvLSTM and its variants are highly effective at modeling both spatial and temporal data, demonstrating strong performance in spatiotemporal prediction tasks. Nevertheless, they can be susceptible to the accumulated prediction error (APE)\cite{wagenmakers2006accumulative, ing2006prediction}.

\subsection{Our Contributions}

With the motivations mentioned above, we propose a novel predictive learning method for realistic V2V channel prediction, focusing on the inherent properties of V2V data and leveraging well-established spatio-temporal predictive learning models. Our key contributions are summarized as follows:

\begin{enumerate}
    \item We address the challenging problem of multi-dimensional V2V channel prediction and introduce a new spatio-temporal predictive learning method. This method incorporates a novel context-conditioned attention mechanism to effectively update spatial and temporal memories within the causal ConvLSTM network. This simple yet effective design leverages the strengths of spatio-temporal predictive learning and intrinsic features to V2V communication systems. 
    \item To enhance the robustness of predictive learning methods across measurements collected from various geometries, we propose a meta-learning framework for training predictive algorithms. This framework effectively addresses the bottleneck issue of APE in RNN-based solutions. Additionally, we incorporate a minor enhancement based on the intrinsic features of V2V data, such as movement status and associated learning difficulty. Our results demonstrate that meta-learning not only applies to but also improves the performance of all predictive learning algorithms. 
    \item We conducted comprehensive case studies to evaluate the performance of various spatiotemporal predictive learning algorithms using measurements from three distinct scenarios, including city canyons and highways. The experimental results show that our proposed method provides accurate and reliable predictions, achieving an average improvement of over 10 dB compared to the baseline method and 3 dB over the state-of-the-art predictive learning method. We will release the data and deep learning algorithm on our research website. \footnote{The dataset and related code will be released on our research website \url{https://wides.usc.edu}  }
\end{enumerate}

The rest of this paper is organized as follows: Section \ref{sec2} introduces the multi-dimensional V2V channel prediction problem and related preliminaries. Section \ref{sec3} elaborates on our proposed method. Section \ref{sec4} provides details on V2V CSI measurement campaigns and evaluations of all spatio-temporal predictive learning algorithms. Finally, Section \ref{sec5} concludes the paper and suggests directions for future research.

\section{V2V Channel Prediction Problem}
\label{sec2}

\subsection{ML-based CSI Prediction Problem Formulation}

The objective of V2V channel prediction is to leverage previously or currently observed CSI to anticipate future channel states. This study concentrates on ML-based approaches to tackle the intricate problem of CSI prediction. Specifically, we employ a neural network, denoted as ${\varphi_{\theta}}$, to address this issue. Given a sequence of CSI frames, ${{{\chi}_1}, \cdots ,{{\chi}_J}}$, the task is to predict a future sequence of length $K$ using ${\varphi_{\theta}}$ based on the $J$ previously observed CSI frames, such that   
\begin{equation} 
\left\{ {{{\chi}_1}, \cdots ,{{\chi}_J}} \right\}\mathop  \Rightarrow \limits^{\varphi_{\theta}} \left\{ {{{\chi}_{J + 1}}, \cdots ,{{\chi}_{J + K}}} \right\}
\label{eq1}  
\end{equation}

In the subsequent analysis, we assume that the lengths of historical and future observations are equivalent. In most V2V systems, including 5G New Radio (NR), transmissions are structured into frames (and possibly further subdivided into subframes and slots). While the frame duration may vary, we use the required prediction times up to 500 milliseconds. Given a "frame-length" of 50 milliseconds, as this corresponds to the periodicity of the signal bursts in our channel measurements. Anticipating the need to predict channels over a period of $500$ ms (e.g., for a longer scheduling horizon in a congested environment), 
this implies that we need to predict up to 10 frames into the future, such as $J=K=10$. To better understand the dynamics of V2V prediction in real-world scenarios, we summarize the key challenges arising from both the measurements and the related prediction methods in the following subsection.

\subsection{Challenges in Multi-Dimensional V2V Channel Prediction}
\subsubsection{The Inherent Properties of the V2V Channel Measurements}
\label{chalegs1}

This subsection details the data structure of the V2V CSI measurements collected using our channel sounder, initially introduced in \cite{wang2017real}. The transmitter (Tx) and receiver (Rx) each consist of 8-element vertically polarized uniform circular dipole arrays mounted on vehicles. During the measurement campaigns, detailed further in Section \ref{mca}, the Tx and Rx communicate over a system with a 5.9 GHz carrier frequency and a 15 MHz bandwidth. As explained in \cite{wang2017real}, our setup consists of a pair of NI-USRP RIOs serving as the main RF transceivers, along with a pair of 8-element switched antenna arrays. The MIMO sounding signal comprises 64 (8 × 8) repetitions of a multi-tone sounding signal, with a total duration $ T_0 $ of 640 $\mu$s. Several guard periods are inserted between the sounding signals to accommodate the settling time of the Tx or Rx switches. The maximum resolvable Doppler shift $ \nu_{\text{max}} $ is given by $ 1/(2T_0) $, approximately 806 Hz, which corresponds to a maximum relative speed of around 148 km/h.  We measure the related MIMO channel burst by burst, each burst containing 30 snapshots, where one snapshot captures the complete MIMO channel, i.e., the transfer function between each Tx and Rx element. Bursts are repeated every 50 ms.   

For $t=1,\cdots, T$, we denote the measured CSI matrices at timestamp (burst) $t$ as ${\bf H}_t \in \mathbb{C}^{ M \times N \times N}$. We account for the burst structure by using variations within a burst to estimate the Doppler spectrum while treating the sequence of bursts as a discrete time series. With these setups, we aim to investigate the time-varying channel in related propagation environments over a spatial-temporal region represented by Time ($T$), Delay ($M$), and Angular ($N \times N$) domains. To better understand the characteristics of our V2V data and its differences from those in the literature, we summarize the datasets used in the context of spatial-temporal prediction in Tab. \ref{stdataset}. 
 
From the perspective of propagation physics, the non-stationary CSI frames in time-varying environments, such as driving through city canyons and highways, create challenging propagation conditions, making it more difficult to derive an effective statistical model \cite{karedal2009geometry, huang2020geometry}. Additionally, we aim to develop a predictive model that can effectively forecast the multi-dimensional CSI frames across four critical domains. However, as illustrated in Table \ref{stdataset}, the structure of our data differs substantially from that of images or videos, making existing predictive learning algorithms potentially less effective for our purposes. Consequently, it is crucial to design specialized models that account for the unique characteristics of our data to enhance prediction performance.

\subsubsection{The Bottleneck Issue in Recurrent-Based Predictive Learning Algorithms}

\begin{table}[!t]
\resizebox{0.495\textwidth}{!}{
\centering
\footnotesize
\begin{tabular}{lcccccccc}
    \hline
    Dataset & Training size & Testing size & Channel & Height & Width & $J$ & $K$ \\
    \hline
    Moving MNIST  & 10,000 & 10,000 & 1 / 3 & 64 & 64 & 10 & 10 \\
    KTH Action & 4,940 & 3,030 & 1 & 128 & 128 & 10 & 20/40 \\
    Human3.6M & 73,404 & 8,582 & 3 & 128 & 128 & 4 & 4 \\
    Kitti \& Caltech & 3,160 & 3,095 & 3 & 128 & 160 & 10 & 1 \\
    TaxiBJ & 20,461 & 500 & 2 & 32 & 32 & 4 & 4 \\
    WeatherBench-M & 54,019 & 2,883 & 4 & 32 & 64 & 4 & 4 \\
    {\bf Ours CSI} & 78,750 & 33,750 & {\textcolor{red}{61}} & 16 & 16 & 10 & 10 \\
    \hline
\end{tabular}
}
\caption{Comparison of the dataset using in the context of spatial-temporal prediction. \label{stdataset}}
\end{table}

In the context of spatiotemporal predictive learning, recurrent-based predictive learning algorithms demonstrate superior performance \cite{zeng2023transformers, tan2023openstl}. For a recurrent-based neural network model, the prediction process is carried out in a recurrent manner, such as:
\begin{equation}
\label{xt1}
{{\hat x}_{t+1}} = {\varphi _\theta }\left( {{x_t},{h^t}} \right),
\end{equation}
where $h^t$ represents the memory state encompassing historical information,  which will be explained in more detail in a later section. The predictive model ${\varphi _\theta }$ corresponds to a neural network
trained to minimize the discrepancy between the predicted future frames and the ground-truth future frames. Given the ground truth ${y_t}, t \in \left\{ {2, \cdots ,J + K} \right\}$, the optimal predictive model is obtained by 
\begin{equation} 
\label{mseloss}
{\theta ^ * } = \mathop {\arg \min }\limits_\theta  \mathcal{L}\left( {{{\hat x}_t},{y_t}} \right)
\end{equation}
where $\mathcal{L}$ denotes a loss function that quantifies the discrepancy.
From the inference shown in Eq. \eqref{mseloss},  the optimization process is performed sequentially, ensuring strong discrepancy capture for the time series.

\begin{figure*}[!ht]
\centering
\includegraphics[width=0.95\textwidth]{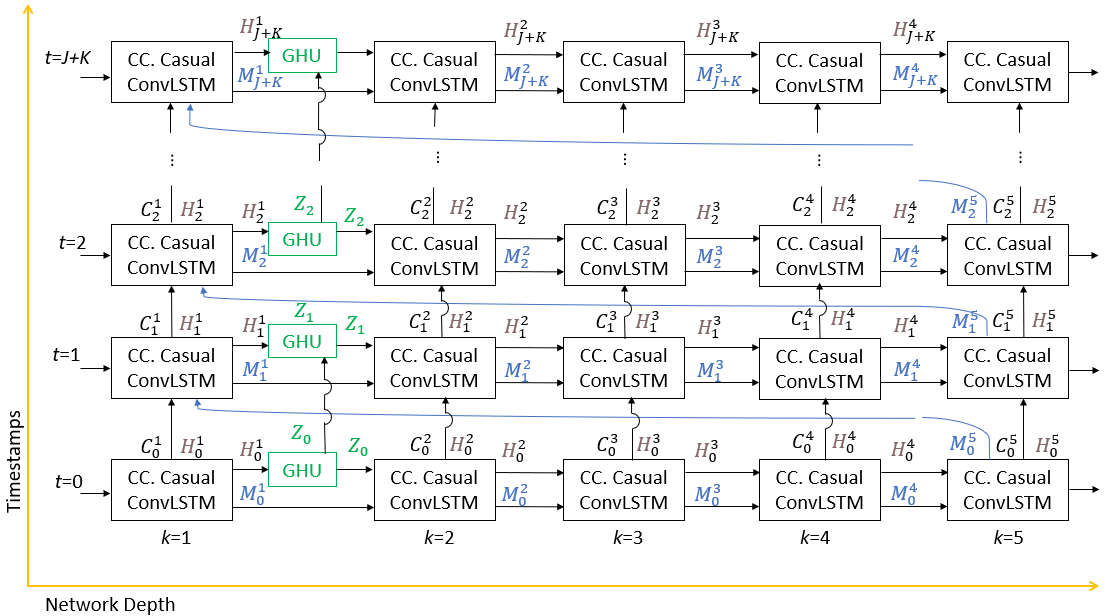} 
\caption{Overall framework of the proposed method. We use the memory attentions as contextual focus. For example, When processing an input sequence, attention mechanisms enable the model to concentrate on various parts of the sequence in a context-sensitive manner. In our model, the temporal context allows the network to learn sequence dependencies in the delay domain, while the spatio-temporal context provides focus in the angular domain. \label{proposedmethod}}
\end{figure*} 

The solutions based on Eq.\eqref{mseloss} account for the dependence within the CSI sequence and are widely adopted in recurrent-based predictive algorithms \cite{tan2023openstl}. To better understand the bottleneck issue, the loss function \eqref{mseloss} can be reformulated to include two components: APE and APE-free, such that:
\begin{equation} 
\label{mse2}
MSE = \underbrace {\sum\limits_{i =  2}^{J} {{{\left( {{\hat x}_t - {y_t}} \right)}^2}} }_{\text{APE-free}} + \underbrace {\sum\limits_{i =  J + 1}^{J+K} {{{\left( {{\hat x}_t - {y_t}} \right)}^2}} }_{\text{APE}}
\end{equation}
The first term represents the APE-free component, as we have ground truth labels from 2 to $ J $. On the other hand, the second term sums the prediction errors from time steps $ J+1 $ to $ J+K $, representing the APE component, as we only have ground truth labels during the training stage but not during the testing stage. 
The bottleneck issue of APE is less pronounced in a supervised learning setup where the number of labeled training samples significantly exceeds that of the unlabeled ones. However, in real-world scenarios, an abundance of labeled training samples cannot be guaranteed. Therefore, it is vital to develop methods that can effectively address APE to ensure robust and reliable channel prediction.

\section{The Proposed Method} 
\label{sec3} 

Building on these motivations, this section details the proposed approach for addressing the complex problem of multi-dimensional CSI prediction. The overall framework is given in Fig. \ref{proposedmethod}. We begin by introducing our predictive learning model, which is tailored to capture dependencies in multi-dimensional CSI sequences effectively. Furthermore, we introduce the network optimization method, which is based on the meta learning framework, incorporating the concept of meta pseudo labels to enhance network training and mitigate the APE bottleneck issue.

\subsection{Preliminaries}

In pursuit of a robust modeling capability that can adaptively handle both short-term and long-term video dependencies in large and highly dynamic datasets, the novel ST-ConvLSTM was proposed, introducing spatiotemporal memory and new recurrent memory updating strategies. Subsequently, a new variant, the CA-ConvLSTM, was developed \cite{wang2018predrnn++} with deep-in-time architectures, further enhancing the network's recurrent depth and representation ability. It demonstrates SOTA performance across multiple datasets, as verified in \cite{tan2023openstl}. Our proposed method follows this recurrent network research line and is built upon the CA-ConvLSTM framework. 

For a better understanding of the proposed method, we first explain the CA-ConvLSTM module, which includes two core parts: The causal ConvLSTM and the gradient highway. A CA-ConvLSTM unit features dual memories: the temporal memory $C^k_t$ and the spatial memory $\mathcal{M}^k_t$. In this notation, the subscript $t$ represents the time step, while the superscript $k$ denotes the $k$-th hidden layer. The current temporal memory $C^k_t$ directly depends on its previous state $C^k_{t-1}$ and is regulated through three gates: a forget gate $f_t$, an input gate $i_t$, and an input modulation gate $g_t$. Meanwhile, the current spatial memory $\mathcal{M}^k_t$ is influenced by $\mathcal{M}^{k-1}_t$ in the deeper transition path. Specifically, for the bottom layer ($k = 1$), the topmost spatial memory at $(t - 1)$ is assigned to $\mathcal{M}^{k-1}_t$. Distinct from the original spatiotemporal LSTM, the CA-ConvLSTM utilizes a cascaded mechanism, where the spatial memory is particularly a function of the temporal memory through an additional set of gate structures. The update equations of the CA-ConvLSTM at the $k$-th layer are as follows: 
\begin{equation}
\label{ca-conv}
\begin{aligned}
\begin{pmatrix}
g_t \\
i_t \\
f_t
\end{pmatrix} &=
\begin{pmatrix}
\tanh \\
\sigma \\
\sigma
\end{pmatrix}
W_1 * \begin{bmatrix}
X_t, H_{t-1}^k, C_{t-1}^k
\end{bmatrix} \quad {(\ref{ca-conv}a}) \\
C_t^k &= f_t \odot C_{t-1}^k + i_t \odot g_t \quad ({\ref{ca-conv}b}) \\
\begin{pmatrix}
g_t' \\
i_t' \\
f_t'
\end{pmatrix} &=
\begin{pmatrix}
\tanh \\
\sigma \\
\sigma
\end{pmatrix}
W_2 * \begin{bmatrix}
X_t, C_t^k, \mathcal{M}_t^{k-1}
\end{bmatrix} \quad ({\ref{ca-conv}c}) \\
\mathcal{M}_t^k &= f_t' \odot \tanh\left( W_3 * \mathcal{M}_t^{k-1} \right) + i_t' \odot g_t' \quad ({\ref{ca-conv}d}) \\
o_t &= \tanh\left( W_4 * \begin{bmatrix}
X_t, C_t^k, \mathcal{M}_t^k
\end{bmatrix} \right) \quad ({\ref{ca-conv}e}) \\
H_t^k &= o_t \odot \tanh\left( W_5 * \begin{bmatrix}
C_t^k, \mathcal{M}_t^k
\end{bmatrix} \right) \quad ({\ref{ca-conv}f})
\end{aligned}
\end{equation}
where $ * $ denotes convolution, $ \odot $ represents element-wise multiplication, $ \sigma $ is the element-wise sigmoid function, square brackets (``[ ]") indicate a concatenation of the tensors, and round brackets denote a system of equations. $ W_1 \sim W_5 $ are convolutional filters, with $ W_3 $ and $ W_5 $ being $ 1 \times 1 $ convolutional filters used for feature fusion while preserving the original dimensions. The final output $ H^k_t $ is determined by both the temporal memory $ C^k_t $ and the spatial memory $ \mathcal{M}^k_t $. 

The CA-ConvLSTM is designed to address the spatiotemporal predictive learning dilemma between deep-in-time structures and vanishing gradients by 1) incorporating a causal LSTM with a cascaded dual memory structure to enhance modeling of short-term dynamics, and 2) integrating a gradient highway unit to provide quick routes for gradients from future predictions to distant past inputs, alleviating the vanishing gradient problem.
Considering the unique data characteristics, as compared and explained in Tab. \ref{stdataset}, it is essential to develop a specialized design that can effectively accommodate these characteristics.

\subsection{On the Proposed Design: Context-conditioned CA-ConvLSTM}

In this work, motivated by underlying characteristics of the V2V CSI data, we present our approach based on the CA-ConvLSTM and integrate context-conditioned attentions (CC. Atten.) to enhance representation ability. 

\subsubsection{Temporal Context}
Our design goal is to develop deep predictive learning models that effectively capture dependencies across multiple domains. To achieve this, we introduce a modulation layer and related feature-wise affine transformation, inspired by \cite{perez2018film, birnbaum2019temporal}, which acts as the context for data in the Delay ($M$) domain (as explained in Section \ref{chalegs1}). Let $X$ be the input, the modulation layer, parameterized by ${W_u}$, is represented as 
\begin{equation}
\label{TAatten0}
\mathbf{s} \in \mathbb{R}^B = \frac{1}{{N \times N}}\sum\limits_{m = 1}^N \sum\limits_{n = 1}^N \mathbf{U},    
\end{equation}
where $\mathbf{U} = {W_u} * X$ represents the transformed input and $\mathbf{s}$ denotes the spatially pooled feature vector. The feature-wise affine transformation is defined by 
\begin{equation} 
\label{TAatten}
TA\left( {X} \right) =  \mathbf{e} \cdot \mathbf{U}, 
\end{equation}
with
\[ \mathbf{e} = \tanh({W_{s1}}\sigma({W_{s2}}\mathbf{s})), \]
where ${W_{s1}}, {W_{s2}}$ are the weights in the affine transformation operator. The symbol $\cdot$ denotes channel-wise multiplication. 

The adaptive channel-wise attention in \eqref{TAatten} provides temporal context for the temporal memory. It starts by aggregating global spatial information using global average pooling (as shown in Eq.\eqref{TAatten0}) to capture Doppler domain statistics. An affine transformation is then applied to learn the significance of each channel, generating weights that recalibrate the original feature maps. This enables the network to adaptively highlight informative features while suppressing less useful ones, enhancing feature discrimination and leading to more efficient and accurate feature learning for our V2V channel data.

\subsubsection{Spatiotemporal Context}

The spatiotemporal context is built on the convolutional block attention module \cite{woo2018cbam}, which is designed to enhance CNNs by sequentially applying frequency domain and spatial domain attention mechanisms to the input feature map $X$. For frequency domain attention, it emphasizes important feature channels using global average pooling and global max pooling, followed by a shared MLP to generate a channel attention map:
\begin{equation}
\mathbf{M}_c(X) = \sigma(\text{MLP}(\text{AvgPool}(X)) + \text{MLP}(\text{MaxPool}(X)))
\end{equation}
This map is then multiplied with the input feature map $X$ to produce the channel-refined feature map $X_c = \mathbf{M}_c(X) \cdot X$. Next, spatial attention highlights significant regions within $X_c$ by pooling along the channel axis using average and max pooling, concatenating these, and applying a convolution layer with a $7 \times 7$ filter to produce a spatial attention map:
\begin{equation}
\mathbf{M}_s(X_c) = \sigma(f^{7 \times 7}([\text{AvgPool}(X_c); \text{MaxPool}(X_c)])) 
\end{equation}
The final output feature map $X_s$ is obtained by multiplying this spatial attention map with the channel-refined feature map:
\begin{equation}
STA(X) = \mathbf{M}_s(X_c) \cdot X_c. 
\end{equation}
The spatiotemporal context is enabled by sequentially applying frequency domain and spatial attention mechanisms. Frequency domain attention focuses on 'what' is important by highlighting significant feature channels using global average and max pooling followed by an MLP. Spatial attention focuses on 'where' is important by emphasizing crucial spatial regions within the feature maps through average and max pooling along the channel axis, concatenation, and a convolutional layer. This dual attention mechanism refines the feature maps, leading to improved feature representation and enhanced performance for our CSI data representation, especially for the spatiotemporal memories in CA-ConvLSTM.

\subsubsection{Context-conditioned CA-ConvLSTM}

With the above explanation of the contexts, we are ready to show the key equation for related context-conditioned memory as follows:
\begin{equation}
\label{mm1}
C_t^k = C_{t - 1}^k + TA\left( {C_{t - 1}^k} \right),
\end{equation}
and
\begin{equation}
\label{mm2}
{\cal M}_{t-1}^k = STA\left( {{\cal M}_{t - 1}^k} \right)
\end{equation}
Compared to the legacy CA-ConvLSTM in Eq. \eqref{ca-conv}, our designs (Eqs. \eqref{mm1} and \eqref{mm2}) incorporate relevant context for both temporal memory ({\ref{ca-conv}b}) and current spatial-temporal memory ({\ref{ca-conv}d}). We keep all other operations the same as in CA-ConvLSTM. For compact math notation, we use the $ContextLSTM$ to denote the proposed network.
Moreover, following \cite{wang2018predrnn++}, the Gradient Highway Unit (GHU) is also adopted to prevent long-term gradients from vanishing quickly. The key equations of the GHU are as follows:
\begin{equation}
\label{ghu}
\begin{aligned}
\mathcal{P}_t &= \tanh(W_{px} * X_t + W_{pz} * Z_{t-1}) \\
\mathcal{S}_t &= \sigma(W_{sx} * X_t + W_{sz} * Z_{t-1}) \\
Z_t &= \mathcal{S}_t \odot \mathcal{P}_t + (1 - \mathcal{S}_t) \odot Z_{t-1}
\end{aligned}
\end{equation}
In \eqref{ghu}, $W_{px}$, $W_{sx}$, $W_{pz}$ and $W_{sz}$ are convolutional filters. The switch gate $S_t$ allows for adaptive learning by balancing the transformed input $P_t$ and the hidden state  $Z_{t-1}$. The GHU is positioned between the first and second causal LSTMs, meaning that here $X_t$ is equal to $H^1_{t}$. This design ensures the preservation of long-term gradients and improves the network's capacity to model complex spatiotemporal dependencies.
In summary, for $k=1,2,...,5$, all the key equations in our context-conditional CA-ConvLSTM can be  expressed as:
\begin{equation}
\label{mm3}
H_t^k,C_t^k,{\cal M}_t^k = ContextLSTM\left( {H_t^{k - 1},H_{t - 1}^k,C_{t - 1}^k,{\cal M}_t^{k - 1}} \right).
\end{equation}
Our designs are inspired by in-context learning, as highlighted in recent surveys and research \cite{dong2023survey, zhang2024makes}. In-context learning significantly enhances the capabilities of large language models (LLMs) by enabling them to perform new tasks purely through inference. This is accomplished by conditioning the model on a few input-label pairs and then making predictions for new inputs based on this contextual information. The proposed method employs two contexts, motivated by research showing that Feature-Wise Modulations can strengthen sequence dependencies \cite{perez2018film, birnbaum2019temporal}. These contexts are designed to provide information for the temporal memory $C^k_t$ and the spatial memory $\mathcal{M}^k_t$. With these dual contexts, the memories in the proposed predictive learning model are updated with minimal correlations, aligning with the memory decoupling mechanism \cite{Wang2023pred}.

\begin{figure*}[]
\centering
\includegraphics[width=0.95\textwidth]{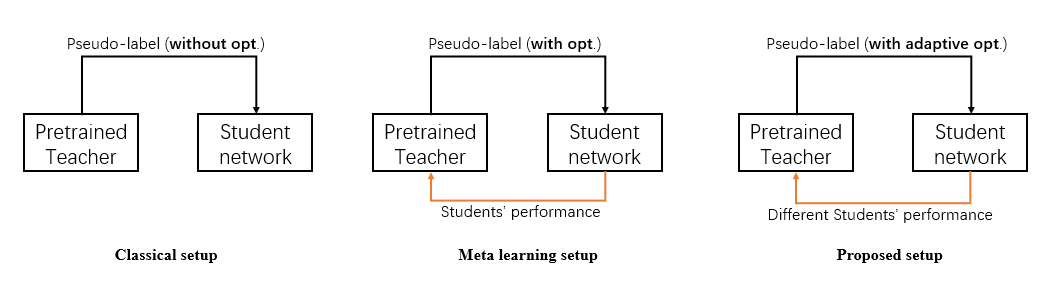} 
\caption{A comparison of different meta learning strategies.\label{fig3}}
\end{figure*} 

\subsection{Model Optimization}
\label{networktrain}

\subsubsection{A Concise Analysis of APE Reduction}

We first provide a concise analysis of APE reduction, explaining the characteristics of APE and outlining possible strategies to address the related challenges. 
 
\begin{remark}
I: The accumulation of APE is unavoidable when utilizing recurrent modules with the MSE function, as defined in Eq. \eqref{mse2}, for neural network training.
\end{remark}

During the training and testing stages of the neural network, as per the problem definition in \eqref{eq1}, the ground truth (GT) CSI frames (the first $J$ frames) are always available, ensuring they are free from APE. However, increased APE is likely to occur when the network is tested across different geometries without information on the remaining $K$ CSI frames that need to be predicted. The prediction error will increase cumulatively when the neural network inference is performed in a recurrent manner. 

\begin{remark}
II: The APE is not significant in the fully supervised training. 
\end{remark}

In the literature on recurrent-based predictive learning, the issue of APE is widespread yet frequently overlooked. Most studies focus on fully supervised training, assuming a significantly larger amount of labeled data compared to unlabeled data. Additionally, it is often assumed that the training and testing data originate from the same distribution. While these assumptions simplify the problem, they do not accurately reflect the complexities encountered in real-world scenarios.

In practice, datasets often exhibit variability and may not follow the same distribution, leading to increased APE when the models are applied to different scenarios or geometries. This oversight in the literature means that many models are not adequately prepared to handle the variability and complexities of diverse datasets. Consequently, APE can accumulate significantly, degrading the model's performance over time and spatial domains. In summary, while recurrent-based predictive learning has made significant strides, addressing the APE issue and the assumption of uniform data distribution is crucial for improving model robustness and performance in diverse real-world scenarios.

\begin{remark}
III: The APE is manageable. 
\end{remark}
On the other hand, the APE can be reduced by enhancing prediction accuracy, as demonstrated by the second term in Eq. \eqref{mse2}. Essentially, any method that improves prediction accuracy can also decrease APE. A straightforward yet effective strategy for this is training the neural network with pseudo labels, as suggested by \cite{lee2013Pseudo}. Pseudo labels involve using the network's own predictions on unlabeled data as additional training data, effectively increasing the amount of training data and helping the model generalize better. By incorporating pseudo labels, the network can iteratively refine its predictions, thus reducing overall prediction error and, consequently, APE. This approach leverages the model's ability to learn from its own outputs, progressively improving performance across both training and testing phases. 

\subsubsection{Meta learning for CSI Prediction:  Pseudo Label Optimization}
\label{Pseudos}

Inspired by meta learning \cite{pham2021meta}, this work introduces meta pseudo labels to reduce the challenging APE errors. The meta learning setup \cite{pham2021meta} employs two types of neural networks: the teacher network ($T$) and the student network ($S$), with parameters denoted by $\theta_T$ and $\theta_S$, respectively. We denote the predictions of the teacher network over the unlabeled CSI $\chi^u$ as $T\left( \theta_T, \chi^u \right)$. Similar definitions apply to the student network, such as $S\left( \theta_S, \chi_l \right)$ and $S\left( \theta_S, \chi^u \right)$. The teacher network teaches the student by minimizing the MSE loss on the unlabeled data: 
\begin{equation}
{{\hat \theta }_S} = \mathop {\arg \min }\limits_{{\theta _S}} {L^u}\left( {{\theta _T},{\theta _S}} \right),
\label{psudoopt} 
\end{equation}
where ${L^u}\left( {{\theta _T},{\theta _S}} \right) = {\mathbb{E}_{{X^u}}}\left[ {MSE\left( {T\left( {{\theta _T};{X^u}} \right),S\left( {{\theta _S};{X^u}} \right)} \right)} \right].$ 

In the meta pseudo labels learning setup, the optimized parameter ${{\hat \theta }_S}$ will be reused in the teacher network optimization, which can be denoted by 
\begin{equation}
\begin{array}{*{20}{c}}
{\mathop {\min }\limits_{{\theta _T}} }&{{L_l}\left( {{{\hat \theta }_S}\left( {{\theta _T}} \right)} \right)}\\
{where}&{{{\hat \theta }_S}\left( {{\theta _T}} \right) = \mathop {\arg \min }\limits_{{\theta _S}} {L^u}\left( {{\theta _T},{\theta _S}} \right)}
\end{array} 
\label{metaopt}  
\end{equation}
We follow the works of \cite{finn2017model, pham2021meta} to solve\footnote{Example code for the regression and supervised experiments is given at
\url{github.com/cbfinn/maml}} the optimization problem in \eqref{metaopt}. The advantageous meta pseudo learning setup not only employs pseudo labels but also incorporates them into the network optimization \cite{finn2017model}, thereby enhancing the performance of both teacher and student networks.

\subsubsection{A Minor Refinement: the Proposed Adaptive Teacher}

The above optimization, as shown in Section \ref{Pseudos}, is based on MSE loss ($L$). Here we provide a minor refinement for it by proposing a weighted MSE loss function, which is given by
\begin{equation}
\label{weighed}
{L_w} = wL,    
\end{equation}
where $\left\langle  \cdot  \right\rangle $ denotes the inner product operator and the weight is defined by 
\[w = \frac{{\exp \left( {{{ - \left\langle {{{\chi}_i},{\bf{\bar \chi}}} \right\rangle } \mathord{\left/
 {\vphantom {{ - \left\langle {{{\chi}_i},{\bf{\bar \chi}}} \right\rangle } 2 }} \right.
 \kern-\nulldelimiterspace} 2 }} \right)}}{{\sum\nolimits_{t = 2}^{J + K} {\exp \left( { - {{\left\langle {{{\chi}_i},{\bf{\bar \chi}}} \right\rangle } \mathord{\left/
 {\vphantom {{\left\langle {{{\chi}_i},{\bf{\bar \chi}}} \right\rangle } 2 }} \right.
 \kern-\nulldelimiterspace} 2 }} \right)} }}\]
$\left\langle  \cdot  \right\rangle $ denotes the inner product operator, and \[{\bf{\bar \chi}} = \frac{1}{{J + K}}\sum\nolimits_i {{{\chi}_i}}. \]
The primary motivation for developing a weighted MSE loss function, where the weights measure the similarity of the input, is to enhance the model's ability to generalize and learn effectively from diverse and complex datasets. Traditional MSE loss functions treat all errors equally, regardless of the similarity between input samples. However, in many real-world scenarios, input data exhibit varying degrees of similarity, and leveraging this information can significantly improve model performance. For example, in our measurement campaign, we observed scenarios where the TX and RX were (nearly) stationary, such as when waiting at a traffic light or stopping at a STOP sign. This led to a high similarity between input samples.

Based on this subsection, we can compare four optimization schemes for predictive learning models. The "Supervised" setup involves fully supervised learning in a same-geo setting. For meta learning in cross-geo settings, as depicted in Fig. \ref{fig3}, three cases are considered: (1) "Without meta learning," where models are trained on a dataset from one scenario and applied to another; (2) "With meta learning," based on standard meta learning \cite{finn2017model} and defined by Eq. \eqref{metaopt}, involving a teacher-student network with identical structures; and (3) "With adaptive meta learning," the proposed setup that also follows Eq. \eqref{metaopt} but incorporates a weighted MSE as defined in Eq. \eqref{weighed}. This latter approach leverages pseudo label optimization and includes a simple yet effective improvement to account for diverse dataset performance during training.

\subsection{Algorithm Implementations}

This subsection details the implementation of the predictive learning algorithm, covering data preprocessing, network parameters, and network training and evaluation.

\subsubsection{Data preprocessing}

Our datasets were collected from three distinct measurement campaigns, with further specifics outlined in Table \ref{campaignTable}. For preprocessing, 
we slice the consecutive CSI MIMO data using a 20-frame-wide, non-overlapping sliding window. Each sequence, therefore, comprises 20 frames in total: 10 frames for input and 10 frames for forecasting. Moreover, we consider a simple approach to convert the raw complex-valued CSI (denoted by $X$) into real values (denoted by $\tilde X$), such that $\tilde X = \left[ {\begin{array}{*{20}{c}}
{{\mathop{\rm Re}\nolimits} \left\{ X \right\}}&{{-\mathop{\rm Im}\nolimits} \left\{ X \right\}}\\
{  {\mathop{\rm Im}\nolimits} \left\{ X \right\}}&{{\mathop{\rm Re}\nolimits} \left\{ X \right\}}
\end{array}} \right] $. We employ antenna-wise normalization for the input and interested readers are referred to existing works for different normalization schemes and input types \cite{o2017introduction, alkhateeb2019deepmimo, chu2023exploiting}.  

\subsubsection{Networks Parameters}

Our predictive learning models incorporate two primary types of networks: recurrent networks and context-conditioned attention. For the recurrent module, 
we have implemented a 5-layer architecture that aims to achieve high prediction quality while maintaining reasonable training time and memory usage. This architecture comprises four CA-ConvLSTM layers with 128, 64, 64, and 64 channels, respectively. On top of the bottom CA-ConvLSTM layer, there is a 128-channel gradient highway unit. Additionally, the convolution filter size is set to 3 for all recurrent units within the architecture. For the neural network defined in the temporal attention mechanism, we utilize a 128-channel convolutional neural network (CNN) for modulation, followed by a feature-wise affine transformation incorporating a global pooling layer, and two multi-layer perceptron (MLP) layers. 

\subsubsection{Network Training and Evaluation}

We evaluate the prediction across various geometries, highlighting the performance of all predictive learning algorithms under both supervised and meta learning frameworks. In the fully supervised learning scheme, the algorithms are trained using data from one geometry and evaluated with data from the same geometry. For example, "S1S1" in the supervised framework refers to both training and evaluation with data from Scenario I. In the meta learning scheme, the models are trained with data from one geometry and evaluated with data from a different geometry. For instance, "S1S2" in the meta learning scheme means training with data from Scenario I and evaluating with data from Scenario II. We use MSE loss for training all models and employ the ADAM optimizer with an initial learning rate of $10^{-3}$. Training is stopped after 10,000 iterations, with a batch size of 8 per iteration, unless otherwise specified. The experiments are implemented in PyTorch and run on a single NVIDIA A100 GPU. 

\begin{table*}[!t]
    \centering
	\begin{tabular}{|l|lll|ll|l|}
		\hline
		\multirow{2}{*}{Campaigns} & \multicolumn{3}{l|}{Datasets}                                                                                                                                                                                                                                                                 & \multicolumn{2}{l|}{Training and Test} & \multirow{2}{*}{Others}                                                                                                                                             \\ \cline{2-6}
		& \multicolumn{1}{l|}{Size (Samples)}                                                                            & \multicolumn{1}{l|}{Environment}                                                               & Mobility                                                                    & \multicolumn{1}{l|}{Same-Geo} & {Cross-Geo}  &                                                                                                                                                                     \\ \hline 
		Scenario I                 & \multicolumn{1}{l|}{\begin{tabular}[c]{@{}l@{}}Training ($\Omega _{Train}^{{S_1}}$): \\ Length: 78,750       \\ Test ($\Omega _{Test}^{{S_1}}$):      \\ Length:  33,750\end{tabular}} & \multicolumn{1}{l|}{\begin{tabular}[c]{@{}l@{}} City \\    \\ And \\ \\ Campus Roads\end{tabular}} & \begin{tabular}[c]{@{}l@{}} MiCW;\\    \\ Low Speed\end{tabular}  & \multicolumn{1}{l|}{\begin{tabular}[c]{@{}l@{}}Supervised;  \\  \\ $\Omega _{Train}^{{S_2}} \Rightarrow \Omega _{Test}^{{S_2}}$ \end{tabular}}           &   {\begin{tabular}[c]{@{}l@{}}Meta;    \\ $\Omega _{Train}^{{S_1}}, \Omega _{Train}^{{S{'}_2}} $ \\ $ \Rightarrow \Omega _{Test}^{{S_2}}$ \end{tabular}}   &  \multirow{3}{*}{\begin{tabular}[c]{@{}l@{}}Data \\ Normalization: \\ Antenna-wise;\\    \\ Sliding \\ Window \\ Length: 10 \\    \\ Data \\ Sample  Size: \\ 10*61*16*16\end{tabular}} \\ \cline{1-6}
		Scenario II                & \multicolumn{1}{l|}{\begin{tabular}[c]{@{}l@{}}Training ($\Omega _{Train}^{{S_2}}$), \\ Length: 80,850     \\ Test ($\Omega _{Test}^{{S_2}}$), \\ Length: 34,650\end{tabular}}            & \multicolumn{1}{l|}{Campus Road}                                                               & \begin{tabular}[c]{@{}l@{}}Moving TX,\\  Static RX;  \\ Low Speed\end{tabular}                                                 & \multicolumn{1}{l|}{\begin{tabular}[c]{@{}l@{}}Supervised;  \\  \\ $\Omega _{Train}^{{S_2}} \Rightarrow \Omega _{Test}^{{S_2}}$ \end{tabular}}           &  {\begin{tabular}[c]{@{}l@{}}Meta;    \\ $\Omega _{Train}^{{S_2}}, \Omega _{Train}^{{S{'}_3}} $ \\ $ \Rightarrow \Omega _{Test}^{{S_3}}$ \end{tabular}}    &                                                                                                                                                                     \\ \cline{1-6}
		Scenario III               & \multicolumn{1}{l|}{\begin{tabular}[c]{@{}l@{}}Training ($\Omega _{Train}^{{S_3}}$), \\ Length: 147,000    \\ Test ($\Omega _{Test}^{{S_3}}$), \\ Length: 63,000\end{tabular}}           & \multicolumn{1}{l|}{Highway Road}                                                              & \begin{tabular}[c]{@{}l@{}}MiCW;\\    \\ High Speed\end{tabular} & \multicolumn{1}{l|}{\begin{tabular}[c]{@{}l@{}}Supervised;  \\  \\ $\Omega _{Train}^{{S_3}} \Rightarrow \Omega _{Test}^{{S_3}}$ \end{tabular}}           &  {\begin{tabular}[c]{@{}l@{}}Meta;    \\ $\Omega _{Train}^{{S_3}}, \Omega _{Train}^{{S{'}_1}} $ \\ $ \Rightarrow \Omega _{Test}^{{S_1}}$ \end{tabular}}    &                                                                                                                                                                     \\ \hline
	\end{tabular}
\caption{Summary of Measurement Campaigns. "MiCW" means TX nad RX were moving in the conveyed way. ${\Omega ^{{S^{'}}}}$ means a subset of ${\Omega ^{{S}}}$.}
\label{campaignTable}
\end{table*}

\section{Case Studies}
 \label{sec4}

In this section, we begin by detailing the data collection process from our three measurement campaigns. We provide a thorough explanation of the movements of TX and RX, the collected CSI measurements, and the basic data preprocessing involved. Following this, we present the experimental results, including comprehensive case studies that evaluate the spatio-temporal predictive learning algorithms across various setups. 

 
\subsection{Measurement Campaign, Performance Metrics and Compared Methods}
\label{mca}

\subsubsection{A brief introduction to the measurement campaign}
\begin{figure*}[!t]
\centering

\subfloat[Scenario I]{ 
\includegraphics[width=0.33\textwidth]{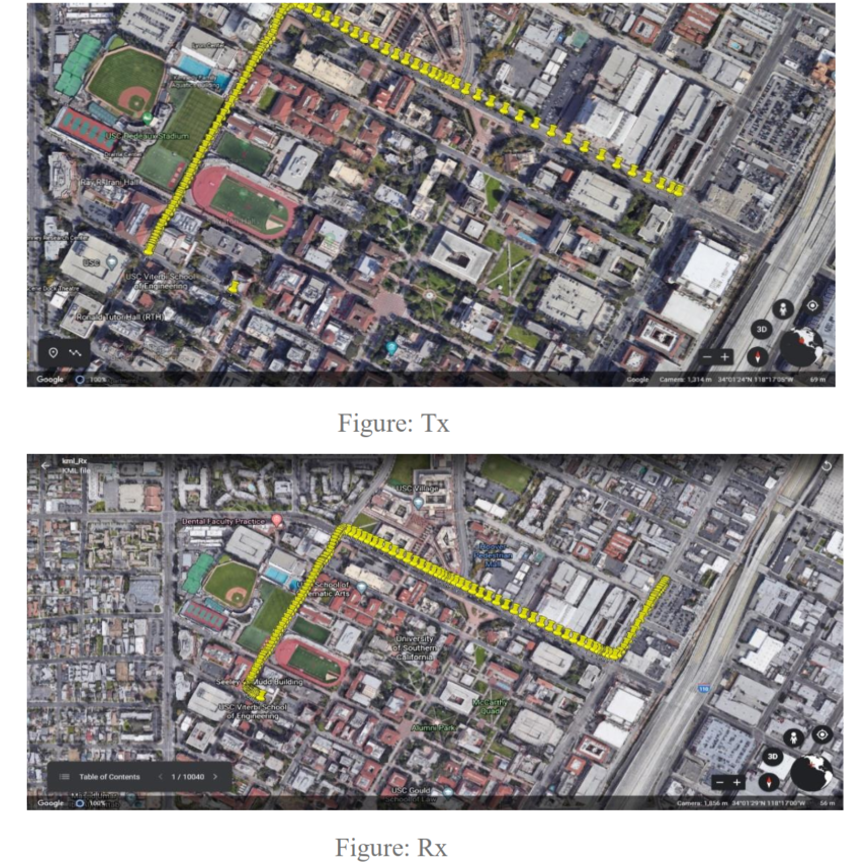}
}
\subfloat[Scenario II]{ 
\includegraphics[width=0.33\textwidth]{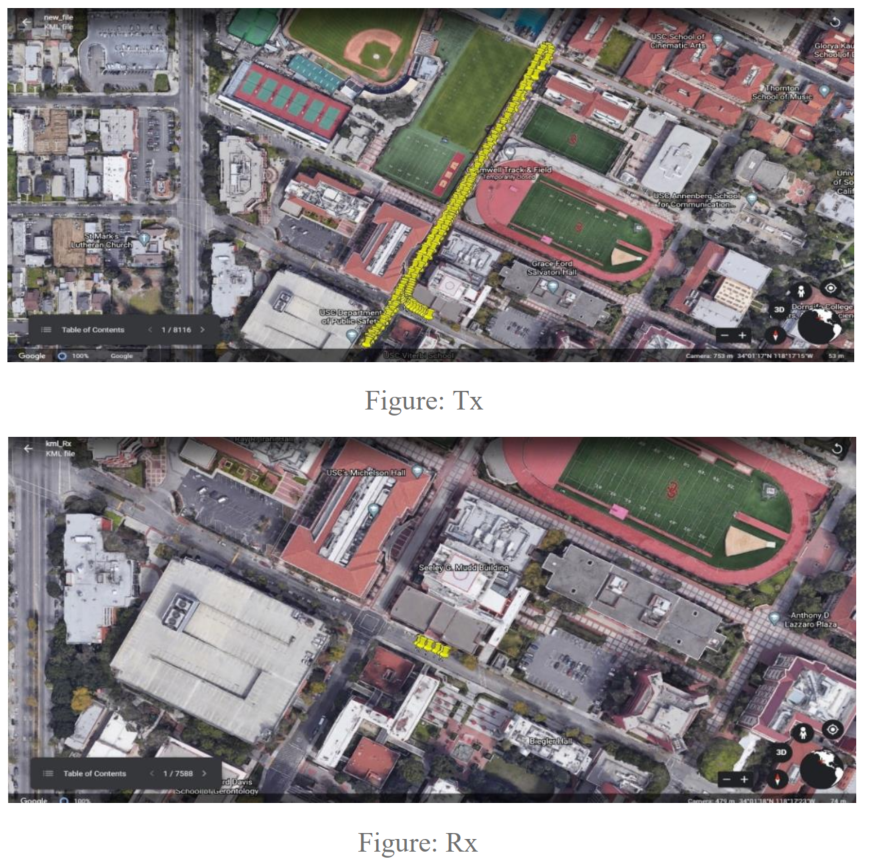}
}
\subfloat[Scenario III]{ 
\includegraphics[width=0.33\textwidth]{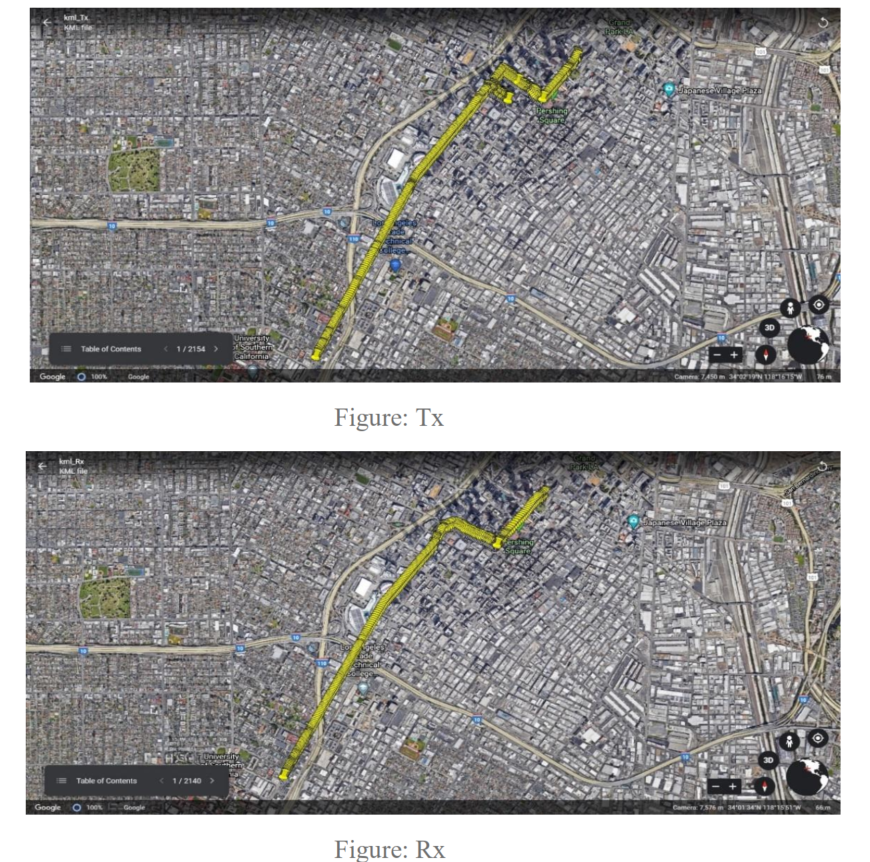}
}
\caption{Measurement campaigns.}
\label{campaigns}
\end{figure*}

In our measurement campaigns, we utilized a channel sounder, as described in Sec. III.B and in more detail in \cite{wang2017real}, to measure the CSI across three challenging scenarios: Scenario I: Mixed City and Campus Road - This scenario combines urban and campus road environments; Scenario II: Campus Road - This scenario focuses solely on the campus road environment; Scenario III: Highway - This scenario pertains to a highway setting. The trajectories of the transmitter (TX) and receiver (RX) for each of these scenarios are depicted in Fig. \ref{campaigns}. In Scenarios I and III, both TX and RX were mobile, moving along predefined paths. Conversely, in Scenario II, the RX remained stationary while the TX was in motion. These configurations were intentionally designed to explore the different mobility patterns in Vehicle-to-Vehicle (V2V) communication systems. A summary of the measured CSI data for each scenario is provided in Tab. \ref{campaignTable}.

\subsubsection{Performance Analysis}
\label{perforeva}

In our evaluation, we employ two different training-test setups. The first setup, referred to as "same geometry training and test" (same-geo), follows the principles of supervised learning. In the same-geo setup, the training and test datasets (CSI measurements) are obtained from the same geometry. We divided the entire dataset into training, validation, and test sets with a ratio of 7:1:2. The second setup, known as "cross-geometry training and test" (cross-geo), involves training and test datasets (CSI measurements) obtained from two different geometries. For instance, we might collect CSI data in Scenario I and evaluate the performance of predictive learning algorithms in Scenario II. 

For our comparison of spatio-temporal predictive learning algorithms, we focus on evaluating the proposed method against two prominent algorithms: ConvLSTM \cite{shi2015convolutional}, and ST-ConvLSTM \cite{Wang2023pred}. ConvLSTM is a well-established algorithm, extensively cited, and has been validated for its effectiveness in CSI prediction problems within wireless communication systems \cite{liu2022spatio}. It utilizes convolutional structures within LSTM units to capture both spatial and temporal dependencies, making it particularly suitable for our evaluation.
ST-ConvLSTM, on the other hand, represents the state-of-the-art in spatio-temporal predictive learning algorithms. ST-ConvLSTM employs an advanced recurrent neural network with novel spatio-temporal memory designs specifically designed to handle complex spatio-temporal dynamics, offering state-of-the-art predictive performance over numerous methods, as verified in \cite{tan2023openstl}. By comparing our proposed method with these two advanced algorithms, we aim to provide a comprehensive assessment of its effectiveness in the context of spatio-temporal predictive learning for CSI prediction.  

Given the ground truth CSI data ${\bf{\chi }}$ and predicted ones ${\bf{\hat \chi }}$, we use normalized mean square error (MSE) and mean absolute error (MAE) as the performance metrics, which are defined as
\begin{equation}
MSE = \frac{{\left\| {\left( {{\bf{\chi }} - {\bf{\hat \chi }}} \right){{\left( {{\bf{\chi }} - {\bf{\hat \chi }}} \right)}^T}} \right\|}}{{\left\| {{\bf{\chi }}{{\bf{\chi }}^T}} \right\|}} \vspace{-0.00001cm},  
\end{equation}
and
\begin{equation}
MAE = \frac{{\sum {\left| {\left( {{\bf{\chi }} - {\bf{\hat \chi }}} \right)} \right|} }}{{\sum {\left| {{\bf{\chi }}} \right|} }}.  
\end{equation}

In our evaluation, we use MSE and MAE to assess the performance of the spatio-temporal predictive learning algorithms. MSE, which measures the average squared difference between predicted and actual values, is sensitive to large errors and provides a smooth optimization landscape, making it useful for variance estimation. MAE, measuring the average absolute difference, is robust to outliers and offers a direct, interpretable measure of average prediction error. By using both metrics, we capture the sensitivity to large errors and the overall robustness of the predictions, ensuring a comprehensive evaluation.

\subsection{Prediction Performance Over Time}

\begin{figure*}[!t]
\centering
\subfloat[Scenario I (MSE)]{ 
\includegraphics[width=0.33\textwidth]{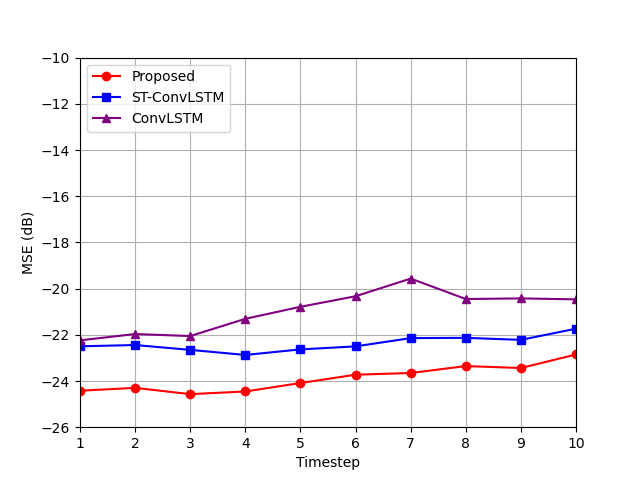}
}
\subfloat[Scenario II (MSE)]{ 
\includegraphics[width=0.33\textwidth]{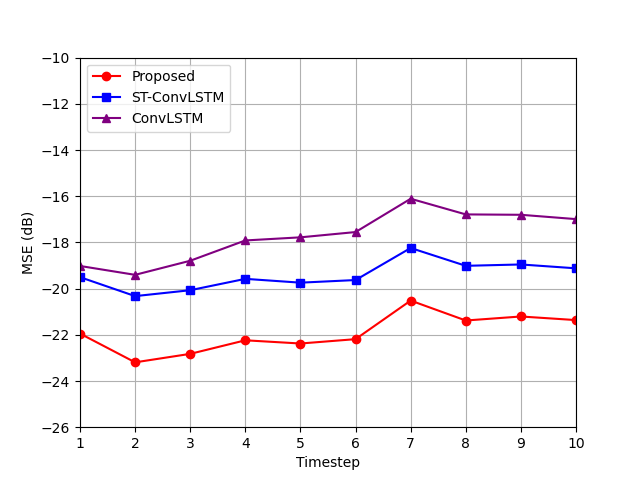}
}
\subfloat[Scenario III (MSE)]{ 
\includegraphics[width=0.33\textwidth]{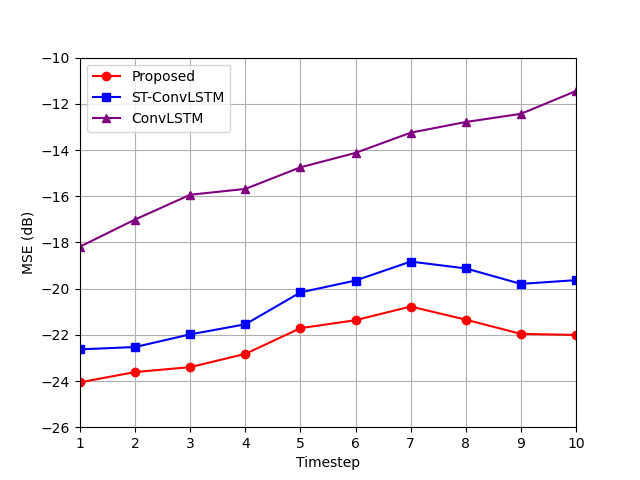}
} \\
\subfloat[Scenario I (MAE)]{ 
\includegraphics[width=0.33\textwidth]{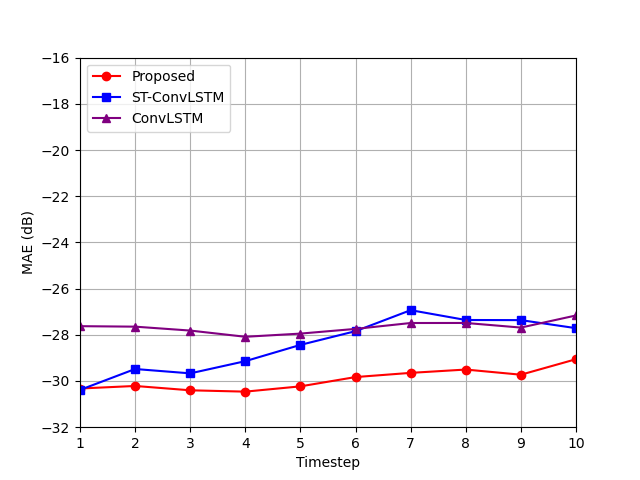}
}
\subfloat[Scenario II (MAE)]{ 
\includegraphics[width=0.33\textwidth]{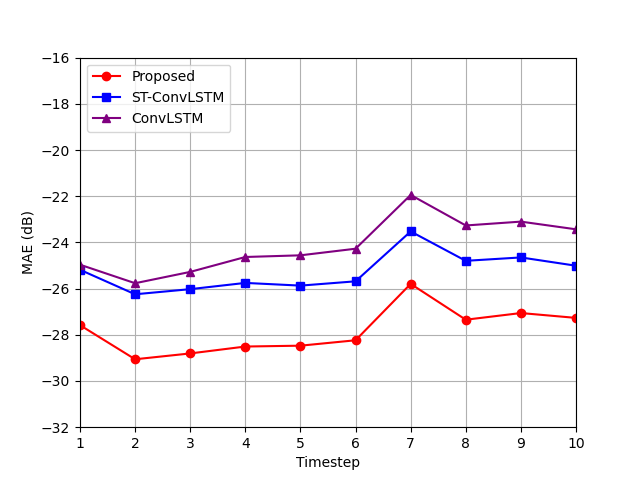}
}
\subfloat[Scenario III (MAE)]{ 
\includegraphics[width=0.33\textwidth]{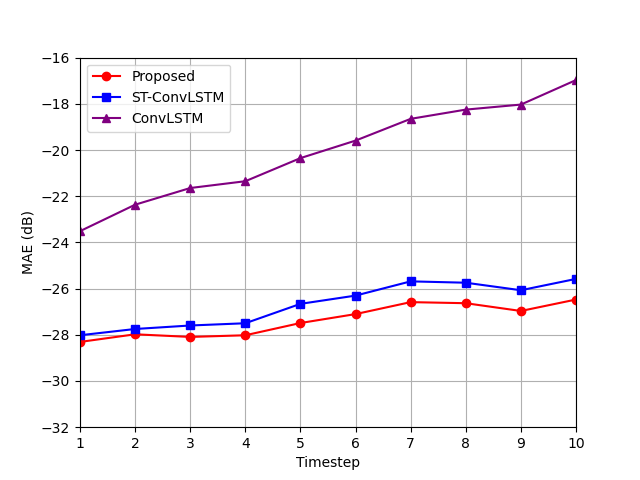}
}
\caption{CSI Prediction performance in three different scenarios, with the upper row displaying MSE results and the lower row showing MAE results.}
\vspace{-0.35cm}
\label{figSAME_GEO_mse}
\end{figure*}

We first present the CSI prediction results over time, using 10 consecutive bursts of CSI MIMO data to predict the next 10 bursts. The corresponding results are shown in Fig. \ref{figSAME_GEO_mse}. The figure illustrates the prediction performance over time for three scenarios, evaluated using MSE and MAE in dB scale.  

For the results related to Scenario I (City and campus roads), shown in the top left and bottom left graphs, the proposed method slightly outperforms ST-ConvLSTM and ConvLSTM, maintaining lower error values across the timesteps. In this low-mobility scenario, both the TX and RX were moving along the USC campus road. Due to the large coherence time, the channel characteristics remain relatively stable  \cite{va2016impact}. As a result, all the predictive learning algorithms exhibit only slight variations in their prediction performances, reflecting the minimal impact of mobility on the channel conditions in this specific context. 

In Scenario II (Campus Roads), the channel undergoes significant changes due to the differing mobility of the TX and RX. The corresponding prediction results indicate that the proposed method maintains superior performance with lower error values; the margin of improvement is more significant than in Scenario I. This underscores the effectiveness of the advanced design of context-conditioned memory updates in spatiotemporal learning algorithms, confirming their capability to handle varying mobility conditions.

In Scenario III (Highway), both the TX and RX were moving in a controlled manner while encountering more environmental dynamics compared to Scenario I, such as the presence of nearby high-speed vehicles. The related prediction results, presented in the top right and bottom right graphs, show the most significant improvement by the proposed method over the other algorithms, particularly as the timesteps increase, with ConvLSTM exhibiting the highest error values. This demonstrates the superiority of spatial and temporal memory designs in both ST-ConvLSTM and our proposed method. Overall, the proposed method shows the best performance in reducing prediction errors (both MSE and MAE) across all scenarios, especially in the more challenging highway environment, underscoring its robustness and effectiveness compared to ST-ConvLSTM and ConvLSTM. Additionally, this proves once again the effectiveness of the advanced design of contested conditioned memory updates. 

\begin{figure*}[!t]
\centering
\subfloat[\label{figSAME_GEOa} Scenario I]{ \label{fig1xxa}
\includegraphics[width=0.33\textwidth]{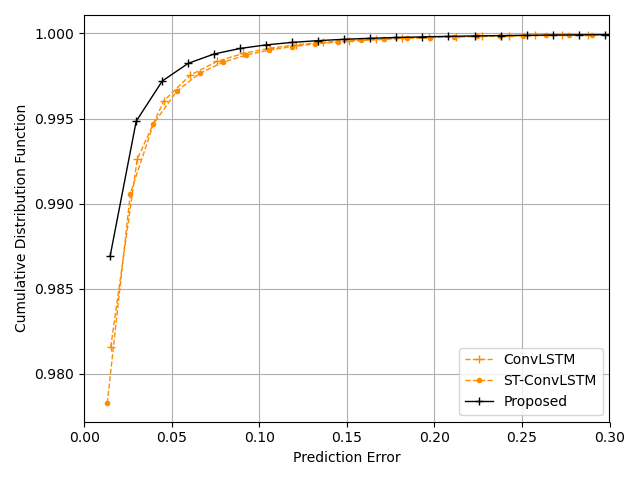}
}
\subfloat[\label{figSAME_GEOb} Scenario II]{ \label{fig1xxb}
\includegraphics[width=0.33\textwidth]{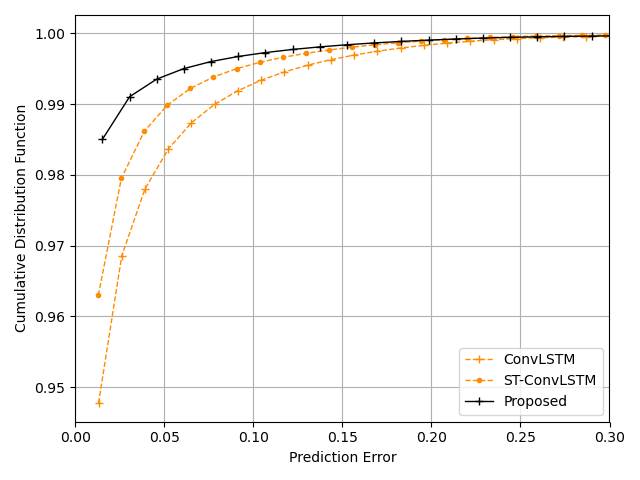}
}
\subfloat[\label{figSAME_GEOc} Scenario III]{ \label{fig1xxc}
\includegraphics[width=0.33\textwidth]{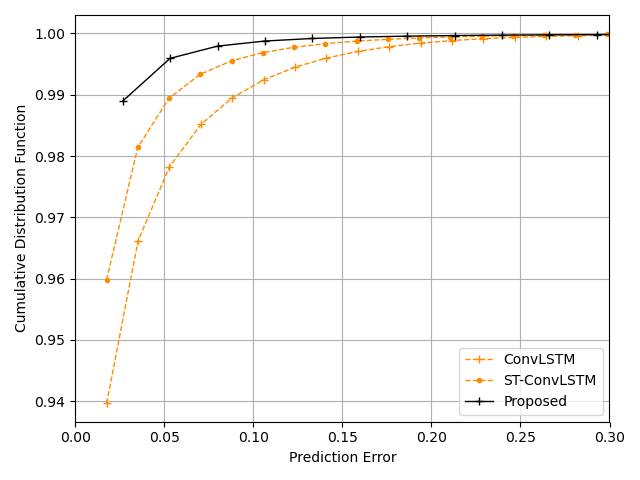}
} 
\caption{The cumulative distribution function of MSEs for all predictive algorithms in three scenarios (same-geo).}
\label{figSAME_GEO}
\end{figure*}
\begin{figure*}[!t]
\centering
\subfloat[\label{figDIF_GEOa} Case I]{  
\includegraphics[width=0.33\textwidth]{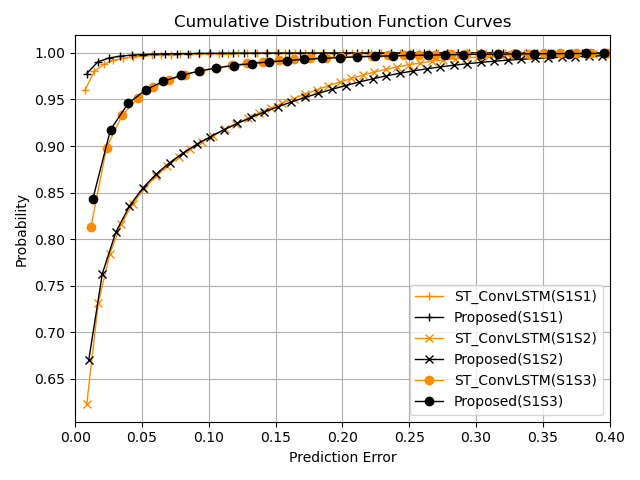}
}
\subfloat[\label{figDIF_GEOb} Case II]{  
\includegraphics[width=0.33\textwidth]{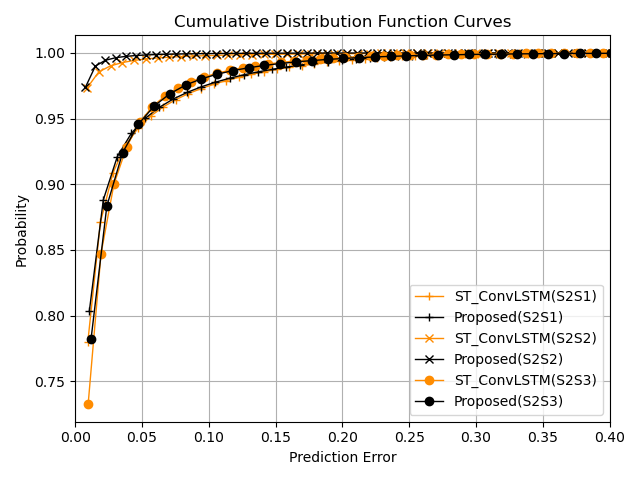}
}
\subfloat[\label{figDIF_GEOc} Case III]{  
\includegraphics[width=0.33\textwidth]{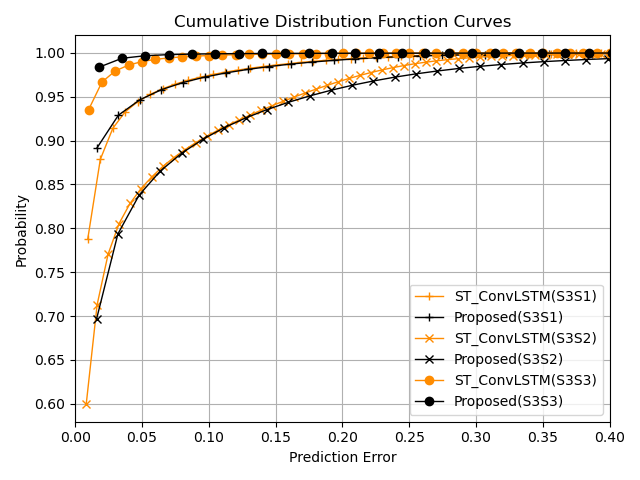}
}
\caption{The cumulative distribution function of MSEs for all predictive algorithms in three scenarios (cross-geo).}
\label{figDIF_GEO}
\end{figure*}

\subsection{Prediction Performance over Different Geometries}

In this subsection, we present the prediction results across various geometries, demonstrating the performance of all employed predictive learning algorithms in both supervised learning and meta learning frameworks. 

\subsubsection{CSI Prediction With Supervised Learning}

We first present results in the same-geo setup, where all predictive learning algorithms were trained in a fully supervised manner using data collected from the same scenarios. We use the cumulative distribution function of MSEs to detail the predictive results in Fig. \ref{figSAME_GEO}. As illustrated in Fig. \ref{figSAME_GEOa} for Scenario I, a low mobility case, all algorithms—ConvLSTM, ST-ConvLSTM, and the Proposed algorithm—demonstrate good and comparable prediction results. This shows that each algorithm effectively captures data variations across various domains of CSI MIMO, including sampling time, bandwidth, and antenna configurations. The comparable results can be attributed to the prediction context, where we forecast ten future bursts of CSI data based on ten historical bursts within a 6.4 ms prediction interval, which can be close to the coherence time. The small data variations within this short time frame result in similar predictive performance across the algorithms, validating their capability to manage the temporal dynamics and spatial characteristics of the CSI MIMO data. Moreover,  in more complex cases (Scenario II and Scenario III), the proposed method significantly outperforms the others, highlighting the benefits of innovative designs such as CC. Atten. in improving spatiotemporal predictive learning performance.

As a comparison, we also present the comprehensive prediction results in a cross-geo setting, where all predictive learning models (ST-ConvLSTM and the proposed method) are trained on data from one scenario but tested on data from a different scenario. Figure \ref{figDIF_GEO} shows that all predictive learning models experience significant performance loss in this challenging setting. This performance degradation is primarily due to the inherent difficulties of V2V channel prediction with measurements collected from different scenarios. Applying trained models from one scenario to data from another proves challenging due to the unique characteristics of CSI data under varying propagation conditions, leading to a severe APE issue. Moreover, we observe that mobility patterns significantly impact prediction performance. For instance, cross-scenario training with similar moving patterns, i.e., S1S3 and S3S1 (Fig. \ref{figDIF_GEO}a and Fig. \ref{figDIF_GEO}c), provides better results than those for different mobility patterns, S2S3 and S2S1 (Fig. \ref{figDIF_GEO}b). This suggests that spatiotemporal predictive learning models can effectively capture the moving patterns, although it remains challenging to apply the learned patterns from one scenario to another. For more accurate and reliable results, we will present the outcomes in the meta learning setting below.
 
\begin{figure}[ht]
\centering
\subfloat[S1-S2]{  
\includegraphics[width=0.45\textwidth]{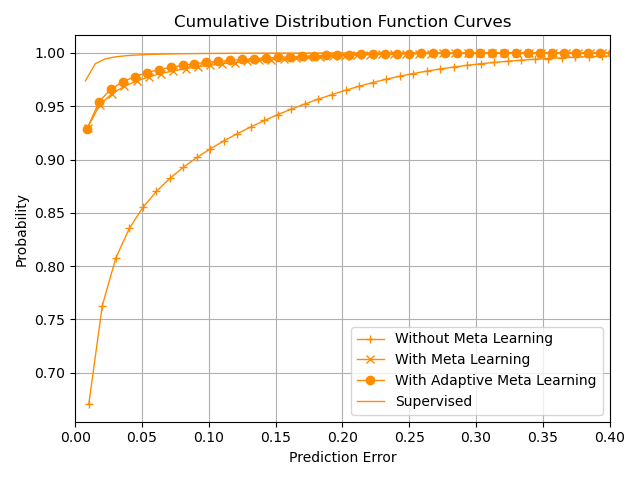}
} \\
\subfloat[S1-S3]{  
\includegraphics[width=0.45\textwidth]{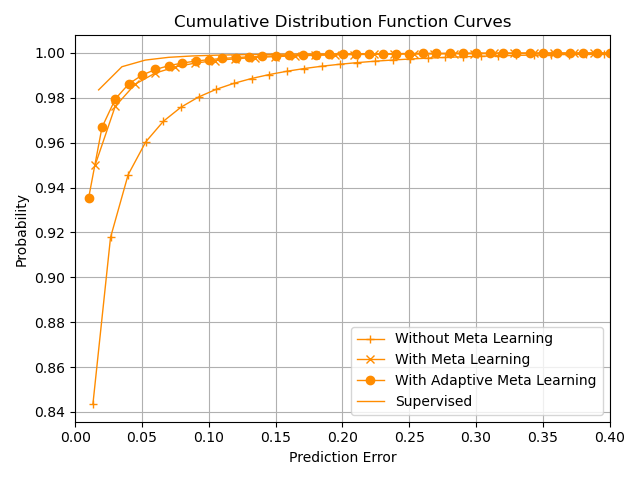}
}
\caption{\label{meta-learn-case}The cumulative distribution function of MSEs for all predictive algorithms in the meta learning framework.}
\end{figure}
\subsubsection{CSI Prediction With meta learning}

We present the prediction results using a cross-geo setup, where the network is trained with labeled data from one scenario and tested with data from a different (unseen) scenario. For the meta learning setup, some unlabeled data from the unseen scenario are also included in the training process. Fig. \ref{meta-learn-case}a corresponds to the scenario S1-S2, while Fig. \ref{meta-learn-case}b corresponds to S1-S3. We use the CDF curves of MSEs for the proposed method within several learning frameworks, including "Without meta learning learning", "With meta learning", and "With adaptive meta learning learning", which have been introduced in Section \ref{networktrain}. For the "Supervised" setup, we consider the fully supervised results (same-geo) as a performance benchmark for the meta learning methods. 

As shown in Fig.\ref{meta-learn-case}, the proposed methods with meta learning schemes, such as "With meta learning" and "With adaptive meta learning," demonstrates superior performance, with "With adaptive meta learning" showing slightly better results due to its adaptive nature. Although there is a performance loss compared to the fully supervised benchmark, this loss is less than the loss between "With meta learning" and "Without meta learning" case. These findings reveal that, in a cross-geo setup, there is considerable potential for performance improvement in predictive learning algorithms, primarily due to the APE bottleneck issue. Meta learning schemes can effectively address this issue by generating and optimizing pseudo labels for unlabeled data, significantly reducing the APE. In summary, the case study in Fig. \ref{meta-learn-case} underscores the importance of leveraging meta learning techniques to enhance prediction accuracy in complex and dynamic scenarios. Additional details, such as the applicability of meta learning for different predictive models and the impact of the number of available labeled samples, will be provided in the following section.

\begin{table*}[ht]
\centering
\small
\begin{tabular}{|l|lll|lll|}
\hline
\multirow{2}{*}{Network Models} & \multicolumn{3}{l|}{MSE(Smoothness)}                                              & \multicolumn{3}{l|}{MAE(Sharpness)}                                               \\ \cline{2-7} 
                                & \multicolumn{1}{l|}{Scenario I} & \multicolumn{1}{l|}{Scenario II} & Scenario III & \multicolumn{1}{l|}{Scenario I} & \multicolumn{1}{l|}{Scenario II} & Scenario III \\ \hline
ConvLSTM                        & \multicolumn{1}{l|}{-20.38}          & \multicolumn{1}{l|}{-17.60}           &      {\bf -14.11}       & \multicolumn{1}{l|}{-27.66}          & \multicolumn{1}{l|}{-24.04}           & {\bf -19.84}            \\ \hline
ST-ConvLSTM                     & \multicolumn{1}{l|}{-22.37}          & \multicolumn{1}{l|}{-19.37}           & -20.38           & \multicolumn{1}{l|}{-28.36}          & \multicolumn{1}{l|}{-25.24}           & -26.64            \\ \hline
CA- ConvLSTM                    & \multicolumn{1}{l|}{-21.65}          & \multicolumn{1}{l|}{-19.66}           & -20.12            & \multicolumn{1}{l|}{-28.04}          & \multicolumn{1}{l|}{-25.42}           & -26.38            \\ \hline
CA- ConvLSTM+ T. Atten.      & \multicolumn{1}{l|}{-22.57}          & \multicolumn{1}{l|}{-20.13}           & -20.55            & \multicolumn{1}{l|}{-28.46}          & \multicolumn{1}{l|}{-25.58}           & -26.44            \\ \hline
CA- ConvLSTM+  S.T. Atten.      & \multicolumn{1}{l|}{-22.61}          & \multicolumn{1}{l|}{-20.29}           & -20.41            & \multicolumn{1}{l|}{-28.62}          & \multicolumn{1}{l|}{-25.64}           & -26.43            \\ \hline
CA- ConvLSTM+ CC. Atten.         & \multicolumn{1}{l|}{-23.66}          & \multicolumn{1}{l|}{-21.49}           & -21.96            & \multicolumn{1}{l|}{-29.76}          & \multicolumn{1}{l|}{-27.48}           & -27.22           \\ \hline
CA- ConvLSTM+ CC. Atten. + GHW        & \multicolumn{1}{l|}{\bf -23.85}          & \multicolumn{1}{l|}{\bf -21.85}           & {\bf -22.18}           & \multicolumn{1}{l|}{\bf -29.92}          & \multicolumn{1}{l|}{\bf -27.76}           & {\bf -27.34}           \\ \hline
\end{tabular}
\caption{\label{tabx1} Ablation studies for each network module in the proposed method.}
\end{table*}
\subsection{The Ablation Studies}
\label{Ablation}
In this section, we present ablation studies that examine the effectiveness of meta learning compared to various predictive learning algorithms, the impact of labeled samples in meta learning, and the role of different network modules in the proposed method. Through these comprehensive studies, we aim to provide readers with a deeper understanding of the key design elements introduced in this work.

\subsubsection{On the impact of labeled samples in meta learning}

This subsection examines the impact of labeled samples in meta learning, using a percentage of available training data with the remainder unlabeled. For comparison, we also present results from supervised learning, where only the same percentage of labeled data is used for training. 
\begin{figure}[!t]
	\centering
	\includegraphics[width=0.99\columnwidth]{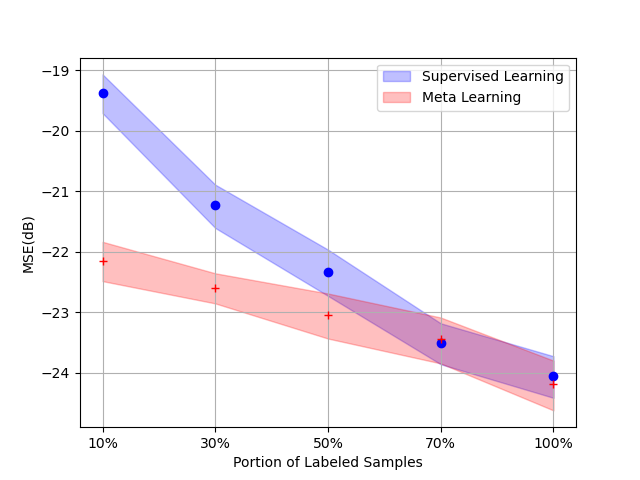} 
	\caption{The Impact of Labeled Samples on meta learning Performance}
	\label{labelsnum}
\end{figure}
Fig. \ref{labelsnum} illustrates the relationship between the portion of labeled samples and the MSE (in dB) for both supervised and meta learning paradigms. As the portion of labeled samples increases from 10$\%$ to 100$\%$, both learning methods exhibit a noticeable decrease in MSE, signifying enhanced performance with more labeled data. Notably, meta learning starts with a significantly lower MSE compared to supervised learning, even with just 10$\%$ of labeled samples, underscoring its initial effectiveness when labeled data is limited. 
When comparing the performance of the two methods, meta learning consistently maintains a lower MSE across all portions of labeled samples. The confidence intervals further reveal that supervised learning shows greater variability at lower portions of labeled samples, indicating less predictable performance.


\subsubsection{The effectiveness of meta learning over different predictive learning algorithms}

\begin{figure}[!t]
	\centering
	\includegraphics[width=0.95\columnwidth]{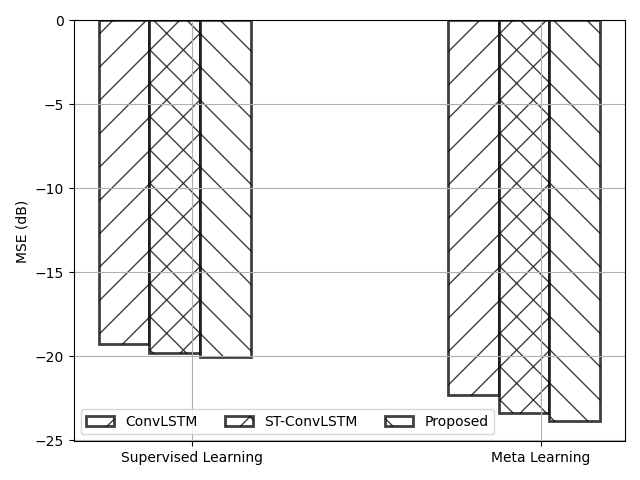} 
	\caption{Comparison of meta learning and Supervised Learning Performance}
	\label{MetaML}
\end{figure}

The bar graph (Fig. \ref{MetaML}) compares the MSE in dB for ConvLSTM, ST-ConvLSTM, and the proposed model under both supervised and meta learning paradigms. To ensure a fair comparison, we used the same quantity of labeled data for network training in both paradigms, specifically ten percent of the available training data from Scenario I. In the meta learning paradigm, the remaining training data was utilized without the associated labelsmeta learning. The test data remained consistent across both paradigms.

As shown in Fig. \ref{MetaML}, the proposed model consistently shows the lowest MSE, indicating superior performance in both learning contexts. All models demonstrate improved performance, but the ST-ConvLSTM and proposed model exhibit the more significant improvement, highlighting its strong adaptability and effectiveness of spatio-temporal memory design within them, especially under meta learning conditions.

\subsubsection{On the role of different network modules in the proposed method}

This subsection examines the role of different network modules in the proposed method. Tab. \ref{tabx1} provides a comprehensive analysis of ablation studies that evaluate the MSE (an indicator of smoothness) and MAE (an indicator of sharpness) across three different scenarios. The comparison includes the baseline ConvLSTM, the state-of-the-art ST-ConvLSTM, and various configurations of the CA-ConvLSTM with different attention mechanisms and additional features. The ConvLSTM model shows the highest MSE and MAE values. In contrast, the ST-ConvLSTM model shows significant improvement, particularly in Scenario III.

The performance of CA-ConvLSTM variants demonstrates further enhancements. Adding Temporal Attention (T. Atten.) and Spatial-Temporal Attention (S.T. Atten.) results in modest improvements over the base CA-ConvLSTM model. However, incorporating CC. Atten. leads to a substantial reduction in both MSE and MAE across all scenarios, showcasing the considerable impact of this attention mechanism on the model's performance. The results highlight the significant role of CC. Atten. in enhancing the model's effectiveness.

The proposed method (CA-ConvLSTM with CC. Atten. and an additional GHU)  emerges as the best-performing model, achieving the lowest MSE and MAE values across all scenarios. This model's superior performance underscores the importance of CC. Atten. and the GHW mechanism in improving both smoothness and sharpness. These enhancements make the proposed method more robust and efficient, significantly outperforming both the baseline ConvLSTM and the state-of-the-art ST-ConvLSTM models. This detailed analysis underscores the effectiveness of the proposed network modules in enhancing predictive performance.

\begin{table}[!t]
\resizebox{0.495\textwidth}{!}{
\begin{tabular}{|ll|llll|}
\hline
\multicolumn{2}{|l|}{\multirow{2}{*}{Methods}}                                                                        & \multicolumn{4}{l|}{Complexity Measure}                                                                                                                                                                                                                                                                                        \\ \cline{3-6} 
\multicolumn{2}{|l|}{}                                                                                                & \multicolumn{1}{l|}{\begin{tabular}[c]{@{}l@{}}Params\\    \\ (M)\end{tabular}} & \multicolumn{1}{l|}{\begin{tabular}[c]{@{}l@{}}FLOPs\\    \\ (G)\end{tabular}} & \multicolumn{1}{l|}{\begin{tabular}[c]{@{}l@{}}Training\\    \\ Time (s)\end{tabular}} & \begin{tabular}[c]{@{}l@{}}Inference\\    \\ Time (s)\end{tabular} \\ \hline
\multicolumn{1}{|l|}{\multirow{3}{*}{\begin{tabular}[c]{@{}l@{}}Supervised    \\ Learning\end{tabular}}} & ConvLSTM & \multicolumn{1}{l|}{\bf 20.34}                                                           & \multicolumn{1}{l|}{\bf 81.19}                                                          & \multicolumn{1}{l|}{\bf 1.916 }                                                                  & {\bf 0.4497}                                                                   \\ \cline{2-6} 
\multicolumn{1}{|l|}{}                                                                                     & ST-ConvLSTM   & \multicolumn{1}{l|}{38.58}                                                           & \multicolumn{1}{l|}{ 171.7   }                                                          & \multicolumn{1}{l|}{2.806}                                                                  &  0.6587                                                                     \\ \cline{2-6} 
\multicolumn{1}{|l|}{}                                                                                     & Proposed & \multicolumn{1}{l|}{39.27}                                                           & \multicolumn{1}{l|}{ 172.0   }                                                          & \multicolumn{1}{l|}{2.812  }                                                                  &             0.6599                                                       \\ \hline
\multicolumn{1}{|l|}{\multirow{3}{*}{\begin{tabular}[c]{@{}l@{}}Meta     \\ Learning\end{tabular}}}                                                       & ConvLSTM & \multicolumn{1}{l|}{40.68}                                                           & \multicolumn{1}{l|}{974.3}                                                          & \multicolumn{1}{l|}{11.91}                                                                  &    0.4497                                                                 \\ \cline{2-6} 
\multicolumn{1}{|l|}{}                                                                                     & ST-ConvLSTM   & \multicolumn{1}{l|}{77.16}                                                           & \multicolumn{1}{l|}{2060}                                                          & \multicolumn{1}{l|}{42.09}                                                                  &  0.6587                                                                    \\ \cline{2-6} 
\multicolumn{1}{|l|}{}                                                                                     & Proposed & \multicolumn{1}{l|}{\bf 78.54}                                                           & \multicolumn{1}{l|}{\bf 2064}                                                          & \multicolumn{1}{l|}{\bf  42.18}                                                                  &     {\bf 0.6599  }                                                              \\ \hline
\end{tabular}
}
\caption{\label{complexitytable} Computational Complexity Comparison.}
\end{table}

\subsection{Computational Complexity Analysis}


Tab. \ref{complexitytable} compares the complexity of ConvLSTM, ST-ConvLSTM, and the proposed model under supervised and meta learning, focusing on parameters,  Floating-point operations per second (FLOPS), training time, and inference time. In supervised learning, ConvLSTM is the most efficient, with the lowest values in all metrics, while ST-ConvLSTM and the proposed model show higher complexity and similar resource usage. Under meta learning, all models become more complex. ConvLSTM's FLOPs and training time increase significantly, reflecting higher computational demands. ST-ConvLSTM's complexity also rises, with more parameters, FLOPs, and longer training time. The proposed model has the highest values in parameters and FLOPs, with slightly longer training and inference times compared to ST-ConvLSTM, making it the most resource-intensive. There is thus, a clear tradeoff between complexity and performance it must be emphasized that the increased complexity mainly impacts training rather than deployment, as indicated by the relatively small (less than $50$ \%) difference consistent inference times across models.



\section{Conclusions}
 \label{sec5}

In this paper, we introduced a novel context-conditioned spatiotemporal predictive model for the challenging V2V channel prediction problem using measurement data. 
We proved the effectiveness of context-conditioned attention, which considers the inherent properties of CSI data, by presenting comprehensive prediction results across various scenarios, ranging from low to high mobility. Moreover, we explained the bottleneck APE issue in recurrent-based predictive models and demonstrated the effectiveness of a meta learning framework in addressing this issue. We verified that meta learning can be applied to all predictive models, albeit at the cost of a higher training budget.

In conclusion, designing a customized spatio-temporal predictive learning algorithm is a complex task, particularly when dealing with the unique characteristics of multiple dimensional CSI data in V2V networks. While state-of-the-art models like ST-ConvLSTM and CA-ConvLSTM provide a foundation, tailored approaches such as context-conditioned attentions are crucial for optimizing performance across diverse scenarios with varying mobilities and propagation environments. Our study demonstrates, for the first time using measurement data, the efficacy of spatio-temporal predictive learning algorithms in different settings. The effectiveness of meta-learning in enhancing network generalization is evident, primarily due to the generation and optimization of pseudo labels, which mitigate cumulative errors in RNN models. This insight is pivotal for improving robustness and accuracy in dynamic environments. We also acknowledge the increased training budget required for superior performance, which, while manageable for most industrial applications, remains an important consideration. We hope this work brings attention to the issue of cumulative errors in RNN families and potential solutions. 

In our future work, we will concentrate on enhancing the efficiency of the meta learning framework, aiming to reduce the computational and time resources required for training. Additionally, we plan to integrate a broader array of sensors to provide richer semantic information about the propagation environment. This integration will introduce new challenges and complexities to the V2V channel prediction problem, ultimately driving further advancements in predictive model accuracy and robustness. 

\section*{Acknowledgment}

The authors thank the proofreading from Omer Serbetci and the valuable discussions with members of the WiDeS group, including Dr. Naveed Abbasi, Dr. Yang Cai, Weiyu Chen, Zihang Cheng, Dr. Hussein Hammoud, Dr. Juhyung Lee, and Dr. Jorge Gomez Ponce. The authors appreciate the support from  Dr. Rui Wang in channel sounder and data prepossessing.

\bibliographystyle{IEEEtran}
\bibliography{sample}

\end{document}